\documentclass[12pt,reqno]{article}

\usepackage{amssymb}
\usepackage{makeidx}

\usepackage[normalem]{ulem}

\makeindex

\def\fnote#1#2{\begingroup\def\thefootnote{#1}\footnote{#2}\addtocounter{footnote}{-1}\endgroup}

\usepackage{hyperref}
\hypersetup{
    colorlinks=true, linkcolor=blue, filecolor=red, urlcolor=cyan, 
    pdftitle={Overleaf Example}
    pdfpagemode=FullScreen,
    }
\usepackage{url}

\usepackage{amssymb}
\usepackage{makeidx}
\usepackage{epsf}
\usepackage{graphicx}        

\usepackage{caption, float}
\usepackage{cite}

\makeindex

\def\inbar{\vrule height1.5ex width.4pt depth0pt}
\def\IB{\relax{\rm I\kern-.18em B}}
\def\IC{\relax\,\hbox{$\inbar\kern-.3em{\rm C}$}}
\def\ID{\relax{\rm I\kern-.18em D}}
\def\IE{\relax{\rm I\kern-.18em E}}
\def\IF{\relax{\rm I\kern-.18em F}}
\def\IG{\relax\,\hbox{$\inbar\kern-.3em{\rm G}$}}
\def\IH{\relax{\rm I\kern-.18em H}}
\def\II{\relax{\rm I\kern-.18em I}}
\def\IK{\relax{\rm I\kern-.18em K}}
\def\IL{\relax{\rm I\kern-.18em L}}
\def\IM{\relax{\rm I\kern-.18em M}}
\def\IN{\relax{\rm I\kern-.18em N}}
\def\IO{\relax\,\hbox{$\inbar\kern-.3em{\rm O}$}}
\def\IP{\relax{\rm I\kern-.18em P}}
\def\IQ{\relax\,\hbox{$\inbar\kern-.3em{\rm Q}$}}
\def\IR{\relax{\rm I\kern-.18em R}}
\def\IT{\relax{\rm I\kern-.18em T}}
\def\ZZ{\relax{\sf Z\kern-.4em Z}}

 \def\kslash{\relax{k\kern-.50em /}}

       \def\g{\gamma}  
\def\e{\epsilon}      
  \def\om{\omega}  \def\Om{\Omega} \def\si{\sigma}

\def\cC{{\cal C}}   \def\cF{{\cal F}}
   
  \def\cM{{\cal M}} 
\def\cO{{\cal O}}   
  \def\cU{{\cal U}}

  \def\mathC{{\mathbb C}}  

    \def\mathP{{\mathbb P}}
\def\mathQ{{\mathbb Q}}
  \def\mathZ{{\mathbb Z}}

     \def\bj{{\bar j}}

\def\bOm{{\bar \Omega}}

\newcommand{\odel}{{\overline \del}}

      \newcommand{\vecnu}{{\vec{\nu}}}

\newcommand{\vecPi}{{\vec{\Pi}}}

\def\fnote#1#2{\begingroup\def\thefootnote{#1}\footnote{#2}\addtocounter{footnote}{-1}\endgroup}

\def\beq{\begin{equation}}
\def\eeq{\end{equation}}
\def\bea{\begin{eqnarray}}
\def\eea{\end{eqnarray}}

\def\lleq#1{\label{#1}\eeq}
\let\nn=\nonumber

\def\notin{\ \hbox{{$\in$}\kern-.51em\hbox{/}}}

\def\ra{{\rightarrow}}

\def\del{\partial}
\def\vphi{\varphi}

  \def\E1Fq{E_1/\IF_q}

\def\rmdeg{{\rm deg}}   \def\rmdet{{\rm det}}      \def\rmdim{{\rm dim}}

   \def\rmord{{\rm ord}}

         \newcommand{\rmcrit}{{\rm crit}}
\newcommand{\rmcs}{{\rm cs}} 
     \def\rmdeg{{\rm deg}} \def\rmdet{{\rm det}}    \def\rmdim{{\rm dim}}
          \newcommand{\rmext}{{\rm ext}}  
                     \def\rmint{{\rm int}}
      \newcommand{\rmks}{{\rm ks}} 
          
\def\rmord{{\rm ord}}      \newcommand{\rmpoly}{{\rm poly}}  
     
\newcommand{\rmpow}{{\rm pow}}  \def\rmprim{{\rm prim}}

\newcommand{\rmPF}{{\rm PF}}

          \newcommand{\rmSpO}{{\rm SpO}}

\def\notdiv{{\relax{~|\kern-.35em /~}}}

\def\boxit#1{
\vbox{\hrule height1pt\hbox{\vrule width1pt\kern0.3cm
\vbox{\kern0.3cm\hbox{$\displaystyle#1$}\kern0.3cm}\kern0.3cm\vrule
width1pt}\hrule height1pt}}

\begin{document}
\parindent=0pt

\baselineskip=17pt
\parskip .1truein
\parindent=0pt

\phantom{what}

\vskip .8truein

\centerline{\Large \bf Special Fano geometry from Feynman integrals}

\vskip .3truein

\centerline{\sc Rolf Schimmrigk\fnote{1}{rschimmr@iu.edu, netahu@yahoo.com}}

\vskip .3truein

\centerline{Dept. of Physics}
\vskip 0.05truein
\centerline{Indiana University South Bend}
\vskip 0.05truein
\centerline{1700 E Mishawaka Ave, South Bend, IN 46615}

\vskip 1truein

\centerline{\bf Abstract}

\begin{quote}
  One of the fundamental open questions in QFT is what kind of functions appear as Feynman integrals.
  In recent years this question has often been considered in a geometric context by interpreting the 
  polynomials that appear in these integrals as defining algebraic varieties. One focal point of the past decade 
  has in particular been the class of Calabi-Yau varieties that arise in some types of Feynman integrals.  
  A class of manifolds that includes CYs as a special case are varieties of special Fano types. These varieties 
  were originally introduced because the class of CY spaces is not closed under mirror symmetry. 
  Their Hodge structure is of a more general type and the middle cohomology in particular is determined 
  by two integers, the dimension of the manifold and a charge $Q$.  In the present paper this class of 
  manifolds is considered in the context of Feynman integrals.
 \end{quote}
  
\renewcommand\thepage{}
\newpage
\parindent=0pt

\pagenumbering{arabic}

 \baselineskip=16pt
 \parskip=0.01truein
 
\tableofcontents

\vskip 0.8truein

\baselineskip=19pt
\parindent=0pt
\parskip=0.15truein

\section{Introduction}

Recent years have seen a renewed effort in the understanding of the nature of higher loop Feynman integrals in terms of both the types 
of functions they generate and in terms of the geometries that they lead to. It has in particular been recognized that Calabi-Yau varieties can 
appear in the parametrized form of the Feynman integrals, which led to a march through the dimensions, starting with elliptic curves, 
continuing with K3 surfaces and CY threefolds, and leading to CY $\ell$-folds for particular sequences of graphs, such as the banana graphs and 
traintrack graphs. While the literature concerned with geometric aspects of Feynman integrals has a long history, and is covered 
in books such as \cite{e66etal, ht66, n71, iz80} (see also  refs. \cite{bek05, am08, k22} for later discussions), 
the focus on CYs is more recent.  References for this discussion include 
    \cite{ b09, bs10, mwz11, mwz12, bv13, bkv14, bkv16, fs18, b18etal, b18etal-b, bf19etal, b19etal, k19etal, vv20, b20etal, f21etal,  
            b21etal, fh22, d22etal,  pww22, dhw22, pww22b, d22etal-b, d23etal, c23etal, g23etal, mh23, dk23etal, m23etal, d24etal, 
            m24etal, j24etal, d24etal-b, fnt24}.  
 Higher dimensional CY spaces associated to higher loop Feynman integals  have recently also been useful in the computation 
 of black hole scattering amplitudes \cite{f23etal, k24etal, f24etal, d24b-etal, f24etal-b}.  
 Calabi-Yau varieties were originally of interest in string compactification \cite{chsw85},  in particular in  the 
context of mirror symmetry \cite{cls89, ls90, gp90}. This framework may seem removed from QFT and  hence it is perhaps initially   
 surprising that they should play a special role in QFT. This raises the question whether there 
are other types of distinguished classes of manifolds that can be identified  in the context of Feynman integrals. 

 In the framework of string theory there are at least two reasons why one might be interested to consider a more general class of varieties 
 that includes CYs as a special case. 
 The first is the existence of exactly solvable models \cite{g87}, and more generally Landau-Ginzburg theories 
 \cite{m88, vw88, lvw89, ks92, krsk92}. 
 Among these string vacua there are models for which a Calabi-Yau compactification is not possible, which raised the question 
 whether such models 
 admit a geometric construction in some other way. This led to the introduction of the class of special Fano varieties which were shown to encode 
 the spectra of the critical string vacua  \cite{rs92, rs94}.  
 A second reason for considering manifolds of special Fano type is that the existence of rigid manifolds presented a difficulty in the 
  context of mirror symmetry of CY manifolds because such rigid spaces seem to preclude the possibility of a mirror.
 This issue also motivates the consideration of the general class of special Fano type geometries because it provides a 
 framework in which mirrors of rigid Calabi-Yau  varieties could be discussed \cite{rs92, rs94}.  The issue of mirror symmetry for rigid CYs was 
 the focus of  ref. \cite{cdp93}, where these special Fano varieties were called generalized Calabi-Yau varieties.  This latter designation
  has in the meantime been used by Hitchin to describe a different class of objects. Since Hitchin's definition has been widely adopted, 
  the present paper continues to use the original name.  A toric analysis of special Fano varieties  
 was discussed in \cite{bb94}, while motivic aspects were considered in \cite{kls08}.  Other aspects of these manifolds were discussed in 
 refs.   \cite{im11a, im11b, fik12}. 
  
   Given the appearance of CYs in Feynman integrals it is natural to ask whether the class of more general geometries given by special 
   Fano varieties is realized in quantum field theory and to what extent they provide a useful geometric framework. 
  It is the purpose of this note to discuss this issue.
  It will become clear below that many Feynman integrals that arise in the standard model lead  to 
Symanzik polynomials that describe varieties of special Fano type, including several infinite sequences of higher loop graphs.
 These manifolds thus provide an interesting framework  for the geometric analysis of Feynman graphs that appear both in particle 
 and gravitational physics.
 
Section two of this paper defines the varieties of special Fano type and discusses some aspects that are relevant in the context of 
Feynman integrals and that were not discussed in the early papers on these manifolds. 
Section three considers these varieties in the context  of graph polynomials and
section four is dedicated to examples. Section five concludes.

\vskip .3truein

\section{Special Fano varieties}

The class of special Fano varieties was originally motivated by the existence of exactly solvable string vacua, based on tensor products of 
$N=2$ supersymmetric minimal CFTs \cite{g87}, and more generally by the framework of Landau-Ginzburg type vacua. 
The construction of these exactly solvable models \cite{ls89, lr88, fkss89}  and LG vacua \cite{ks92, krsk92} led to  models with  spectra 
that contain modes corresponding to complex deformations but that do not have modes corresponding 
to  K\"ahler deformations.\fnote{1}{The list of models can be found for example in the CERN library \cite{cern-prep}.} 
 Such models are candidates for  mirror theories of rigid Calabi-Yau string vacua and the question arose what kind of geometry 
 could be associated to such 
exactly solvable models and their Landau-Ginzburg mean field theories.  The proposal in \cite{rs92} is based on a single 
scaling relation that leads to varieties $X$ that are in general higher-dimensional spaces of Fano type, rather than manifolds of CY type.
As such these spaces have a first Chern class $c_1(X)$ that is positive instead of vanishing, as in the CY case. For the class 
considered in \cite{rs92} the manifolds are however not  of general Fano type but are defined by a quantization condition that 
generalizes the vanishing constraint that defines CY spaces.  

It is easiest to consider Fano varieties of special type in the context of weighted projective hypersurfaces  embedded in 
a weighted projective space and this will suffice for the present paper. In the notation adopted in \cite{cls89} the projective space 
is denoted in terms of the weights
 $w_i$ as $\mathP_{(w_1,....,w_{n+2})}$ and the notation $\mathP_{(w_1,....,w_{n+2})}[d]$ is used to denote the set of 
 varieties $X_n$ of degree $d$ and dimension $n$. 
When all the weights are unity this configuration space 
 is abbreviated as $\mathP_{n+1}[d]$, following the notation of \cite{cdls87}.
In the context of vacua of critical string theory and the issue of mirror manifolds the cohomology induced by the ambient projective 
space does not contribute to the critical spectrum. Hence even though these spaces are K\"ahler, the ambient contribution to the 
 K\"ahler sector is not critical.

The quantization condition for the first Chern class of Fano varieties of special type can be formulated as
\beq
 c_1(X) =~ (Q-1) dh,
 \eeq
 where $Q$ is an integer which has an interpretation as a Landau-Ginzburg charge, and $h$ is the pull-back of the hyperplane class 
 of the ambient space. In terms of the weights $w_i$ this takes the form 
 \beq
  \sum_{i=1}^{n+2} w_i ~=~ Qd.
  \eeq
   For $Q=1$ this  recovers the class of weighted CY hypersurfaces. 
 It was shown in \cite{rs92, rs94} that  special Fano varieties are closely related to string vacua in the critical dimension $D_\rmcrit$ related to 
 the dimension $n$ of the special Fano manifold as   $D_\rmcrit = (n -2(Q-1))$ and that they in particular provide the complete spectrum of the 
 critical dimension vacua. 
 
 One of the characteristic features of special Fano varieties is that the Hodge decomposition of the  intermediate cohomology 
 of $X_n \in \mathP_{(w_1,...,w_{n+2})}[d]$ takes a form that depends on the charge $Q$ as
  \beq
  H^n(X_n) = \bigoplus_{i=0}^{n-2(Q-1)} H^{n-(Q-1)-i, (Q-1)+i }(X_n).
 \lleq{hodge-decomp}
 The  corresponding Hodge numbers $h^{p,q} = \rmdim H^{p,q}(X)$ are symmetric because these spaces are of K\"ahler type. 
  When $Q=1$ this reduces to the case of CYs which have cohomology of degree $(n,0)$, while for $Q>1$ the first nontrivial group that appears 
  in this Hodge decomposition  is shifted  inward in the intermediate part of the Hodge diamond. This arises from the fact that the differential 
  form $\Om_Q$ is given by the integral
  \beq
  \Om_Q ~=~ \frac{1}{2\pi i}   \oint_C ~\frac{\si}{p^Q}
   \lleq{Omega-integral}
   where $p$ is the defining polynomial, $\si$ is the projective measure
   \beq
   \si ~=~ \sum_{i=1}^{n+2} (-1)^{i-1} w_i x_i dx_1 \wedge \cdots \wedge \widehat{dx_i} \wedge \cdots \wedge dx_{n+2}
   \eeq
   and $C$ is a closed curve that encircles the zero locus of $p$. Here the hat indicates deletion of the corresponding factor.
    
    The quantization condition of the first Chern class  guarantees that the differential form $\si/p^Q$ is invariant under projective scaling.  
    Strictly speaking, the residue above takes values in the Hodge filtration at level $F^{{Q-1}} H^n(X_n)$, where the filtration is defined as
    (see \cite{g69}; background references in algebraic geometry include \cite{gh78, v02, cmp03, h23, ps08})
    \beq
     F^r H^n(X_n) =~ \bigoplus_{j=0}^r H^{n-j, j}(X_n),
     \eeq
     but because the variety is of special Fano type the image lands in $H^{n-(Q-1), (Q-1)}(X_n)$. The remaining elements of 
     the intermediate cohomology 
   of the smooth fibers are obtained via the  multiplication of polynomials $q(x_i)$ of appropriate degree, leading to differential forms on 
   the complement of $X_n$ of the form
   \beq
   \om_r ~=~ \frac{q \si}{p^{Q+r}}, ~~~~\rmdeg ~q ~=~ (Q+r) d  ~-~ (n+2).
   \eeq
   The residue image of these forms takes values in the primitive part of the sector $F^{Q+r-1}H^n(X_n)$ of the Hodge filtration defined above.
   This primitive part is given by the kernel of the map defined on $H^q(X, \mathQ)$ by the product with $\om^{n-q+1}$  for 
   $\om\in H^2(X)$ induced by the ambient space form. 
   While this primitive part in general is only a subspace of the cohomology, the full group can be recovered via the Lefschetz decomposition as
   \beq
   H^q(X, \mathQ) ~=~ \bigoplus_r L^r H^{q-2r}(X, \mathQ)_\rmprim,
   \eeq
   where $L$ denotes the multiplication by $\om$.
   The cohomology $H^m(X_n)$  with degrees $m<n$ is determined for projective hypersurfaces by the cohomology of the ambient space and 
   for weighted spaces gets contributions from the singularities.
   
  The form $\Om_Q$ allows the identification of the cohomology group in the Hodge decomposition that parametrizes the deformations of the complex structure. 
  It follows from the Maurer-Cartan integrability equation that the first order deformation of the complex structure is described by 
  the  tangent bundle cohomology $H^{0,1}_\odel(X,T_X)$, where  $T_X$ is the holomorphic part  tangent bundle. 
   For special Fano varieties  $X_n^Q$ of complex dimension  $n$ and charge $Q$ the map  
   is given by 
  \beq
   H^{0,1}_\odel (X_n^Q, T_{X_n^Q}) ~\cong ~ H^{n-Q, Q}(X_n^Q).
   \eeq
   The complex deformations of special Fano manifolds can be shown to be unobstructed via  vanishing theorems of Kodaira as well as 
   Akizuki and Nakano \cite{an54}.

   The relation between the cohomology of the special Fano variety and the lower dimensional critical manifolds shows that 
   certain motives of the former determine motives of the latter. This was analyzed in some detail in \cite{kls08}, 
   where motives associated to sub-sectors of the intermediate cohomology
   $H^n(X_n)$ of the special Fano varieties were shown to be isomorphic to the corresponding motives derived from 
   the intermediate cohomology of the critical manifold.  The examples considered there include 
   special Fano varieties in configurations that appear also in Feynman integrals, examples of 
   which will be discussed in a later section below. 
   The presence of the charge $Q$ thus  decouples, in the sense described above, the dimension of the variety 
  determined from the Feynman graph from the cohomology. 
  
  The decoupling just described between the width of the middle cohomology from the dimension of the variety has 
  implications for the group generated 
  by transformations $T$ that are induced on the cohomology when solutions of the period equation are transported along 
  closed paths that encircle the singular points of the differential equation. The main original result concerning the structure of $T$ is the 
  monodromy theorem of Landman \cite{l66, l73}.
  This result states that for a variety of dimension $n$ the matrices $T$ that represent the monodromy action on the 
  middle cohomology are quasi-unipotent,
  with the exponent of nilpotence determined by the degree of the cohomology as $( T^N - {\bf 1})^{n+1} = 0$ for some integer $N$.
   This formulation is adequate for CYs, see e.g. the discussions in \cite{ck99}  and \cite{b21etal}, because the width of the 
   intermediate cohomology of such 
   varieties directly correlates with the dimension of the variety. In general  however the exponent that gives the degree 
   of nilpotence is given 
   by the width of the Hodge decomposition. This was conjectured by Griffiths \cite{g69} and proven later 
   by Katz \cite{k70}. Hence the monodromy theorem takes the stronger form 
   \beq
   \left( T^N ~-~ {\bf 1}\right)^{w_n} ~=~ 0,
   \eeq
   where $w_n$ now denotes the width of the Hodge decomposition of the intermediate cohomology of the variety. 
   For special Fano varieties
   the exponent is in particular  determined    not only by the dimension of the variety but also by its charge $Q$. 
   This has an impact on the log degree of the 
   solutions of Feynman graph differential equations of special Fano type.
   
  The variation of the cohomology in a family of varieties leads in turn to the variation of the periods of the variety. 
  The cohomology class $\Om_Q$ for 
  special Fano varieties in particular can be integrated  against cycles $\g_i \in H_n(X_n)$, leading to
  \beq
   \Pi_i ~=~ \int_{\g_i} \Om,
   \lleq{periods-Om}
   and more generally for the other forms in $H^n(X_n)$.
  The structure of the periods can be described in terms of the Picard-Fuchs equation \cite{g66a}.
The order of this differential operator is determined by the Hodge structure of the intermediate cohomology, 
which for manifolds $X_n$ of CY type is directly given via the dimension $n$ as $(n+1)$. 
In the case of special Fano manifolds the charge $Q$ 
enters as a second characteristic that determines 
the cohomology.  
One of the key characteristics of special Fano varieties is that  the Hodge decomposition (\ref{hodge-decomp}) 
 thus shows that for families 
$X_n(t_i)$ the order of the Picard-Fuchs differential operator $D_\rmPF$ is affected by the charge $Q$. 
In simple cases it is reduced compared to that of Calabi-Yau varieties, with an order given by
  $\rmord~ D_\rmPF = (n+3-2Q)$.
  Hence, while Picard-Fuchs equations of special Fano varieties  can be obtained in a way similar to the case of CYs, 
   the order of the period differential equation is no longer given directly by the dimension of the variety. 
 For the cubic sevenfold $X_7 \in \mathP_8[3]$  this was considered in ref.  \cite{cdp93}. 
For singular varieties the motive changes its rank, as illustrated in \cite{kls10} in the context of CY varieties. As a consequence
the order of the associated differential equation changes as well.

The moduli space of complex deformations $\cM_\rmcs$ 
of special Fano varieties described above are of K\"ahler type, with the K\"ahler potential $K_\rmcs$ 
determined by the form $\Om_Q$ as
\beq
 K_\rmcs~=~ -\ln ~  i^{n^2 -2(Q-1)}  \int_X \Om_Q \wedge \bOm_Q.
 \eeq
 By expanding $\Om_Q$ in terms of a basis $\om^i \in H^n(X_n)$ that is dual to a homology basis $\g_i \in H_n(X_n, \mathZ)$ 
 with intersection form $\cC_{ij} = \g_i\cdot \g_j$ this K\"ahler potential can be written in terms of the periods as
 \beq
  K_\rmcs ~=~ - \ln ~ i^{n^2 -2(Q-1)} ~\vecPi^\dagger ~\cC ~ \vecPi.
  \lleq{kaehler-pot-cs2} 
  This potential then  determines the Weil-Petersson metric $G_{i\bj}= \del_i \del_\bj K_\rmcs$ on the moduli space of complex deformations. 
  For the K\"ahler deformations the K\"ahler 
 geometry is determined by the K\"ahler form $J$ on the special Fano manifolds, with associated K\"ahler potential 
 \beq
  K_\rmks ~=~ \ln  \int_X J^n
  \eeq
  and corresponding K\"ahler metric. 
 
\vskip .3truein

\section{Special Fano varieties in QFT}

For a Feynman graph with $\ell$ loops, $n_\rmint$ internal lines and $n_\rmext$ external momenta $p_i$ 
the lattice of associated integrals in the lattice $\vecnu \in \mathZ^{n_\rmint}$ can be written as
\beq
 I_\vecnu(p_i, m_a)  ~=~  \int \prod_{j=1}^\ell [d^Dk_j]   \prod_{a=1}^{n_\rmint} \frac{1}{D_a^{\nu_a}(q_a, m_a)},
 \eeq
 where $[d^Dk_j]$ is the measure associated to the independent loop momenta $k_j$, the details of which depend on the adopted conventions.
  $D_a(q_a, m_a)$ are the propagators of the internal lines, with momentum flows $q_a(k_j, p_i)$ constrained by momentum conservation 
  in the $n_v =(n_\rmint +1 -\ell)$ vertices.  The exponents $\nu_a$ are collected in the vector $\nu_a$. 
  The parametric representation of such integrals in terms of the Feynman variables $z_a$  can be written in the following form
   (see for example \cite{e66etal, iz80, w22})
\beq
  I_\vecnu(p_i, m_a) ~=~ c(\nu_a, D)
      \int_{x_a\geq 0} \left(\prod_{a=1}^{n_\rmint} z_a^{\nu_a-1} dz_a \right)
    ~\delta\left(1- \sum_{a=1}^{n_\rmint} z_a\right) ~
      ~\frac{\cU(z_a)^{\nu - \frac{(\ell+1) D}{2}}}{\cF(z_a)^{\nu - \frac{\ell D}{2}}} ,
 \eeq
 where $\nu = \sum_a \nu_a$ and $c(\nu_a, D)$ is a constant obtained in the process that will not play a role in the following. 
 Here the first and second Symanzik  polynomials $\cU$ and $\cF$ are obtained via the expansion  
 \beq
  \sum_a x_a (q_a^2 -m^2) ~=~ M_{j_1j_2} k_{j_1}k_{j_2} ~+~ 2Q_j k_j + J
  \lleq{expansion}
   in terms of the matrix  $N=(\rmdet~M)M^{-1}$ as
  \bea
  \cU &=& \rmdet ~M \nn \\
   \cF &=& (\rmdet~M)J ~-~ Q^t N Q.
  \eea
 The degrees of the Symanzik polynomials are determined by the loop number as 
 \bea
 \rmdeg ~\cU &=& \ell \nn \\
 \rmdeg~ \cF &=& \ell+1,
 \eea
 respectively. The polynomial $\cF$ depends on the physical parameters given by the invariants $p_i\cdot p_j$ of the external momenta and 
 the masses $m_a^2$. It therefore defines a multiparameter deformation family as a configuration in the sense introduced in the previous section. 
 In the projective case these polynomials thus lead to hypersurfaces $X_\cU$ and $X_\cF$ that are elements in the following configurations
 \beq
  X_\cU ~\in ~ \mathP_{n_\rmint-1}[\ell], ~~~~~~X_\cF ~\in ~ \mathP_{n_\rmint-1} [\ell+1].
  \eeq
 The varieties $X_\cU$ and $X_\cF$ are of special Fano type if for some integer $Q$ the configurations satisfy 
 $ Q\ell = n_\rmint$ or $Q(\ell+1)= n_\rmint$, respectively. The corresponding charges of the Feynman integral are denoted 
 by $Q_\cU$ and $Q_\cF$.
 
 Feynman integrals with $\nu_a=1$ for all $a$  play a distinguished role in the lattice of Feynman integrals. 
  For such graphs the special Fano quantization condition leads to the exponents of the Symanzik quotient as
 \bea
  \rmpow_\cU &=& Q_\cU\ell ~-~ \frac{(\ell+1)D}{2} \nn \\
  \rmpow_\cF &=&  Q_\cF (\ell+1)- \frac{\ell D}{2}.
 \eea
A simplification often considered in the literature is the case when the first Symanzik polynomial does not appear in the Feynman representation 
of the integral. In this case the focus is on the family of varieties $X_\cF$ that contains the dependence on the 
kinematic parameters and the masses.  This occurs for example for the $\ell$-loop banana graphs when spacetime is two-dimensional. 
 Imposing such a vanishing constraint for the $\cU$ exponent  in the context of special Fano type integrals leads to the exponent of the 
 second Symanzik polynomial as 
\beq
 \rmpow_\cF ~=~ \frac{1}{2Q_\cU - D} \left(2Q_\cU Q_\cF - \frac{D^2}{2}\right).
 \lleq{pow-F}
 This relation leads to cases in four dimensions where the exponent reduces to $\rmpow_\cF = Q_\cF$, which recovers the integrand 
 considered above in eq. (\ref{Omega-integral}). Such Feynman graphs are for example relevant for standard model physics,
 as will be discussed in the next section.
 
There are many Feynman graphs for which either $X_\cF$ or $X_\cU$, or both, are of special Fano type. 
 Prominent examples include higher loop graphs in the cubic scalar theory that are obtained after tensor-spinor reductions 
 of graphs in the standard model.
 There exist also infinite sequences of graphs with increasing number of loops that lead to 
 special Fano varieties.  Examples of graphs that lead to special Fano type varieties are discussed in the next section.

 It was emphasized in \cite{pt16, pt17} that maximally cut Feynman integrals provide a useful tool in the theory, in particular in 
 the context of  solutions of the differential equations obtained via IBP relations of the master integrals. 
 It was subsequently realized in ref. \cite{v18} that such maximally cut
 integrals can be expressed in terms of the Symanzik polynomials like the Feynman integrals but with the distinction of a 
 compact integration 
 domain without boundary, given by tori $T$.  In the context of Feynman integrals  of special Fano type we can consider graphs 
 for which the exponent of the first Symanzik polynomial vanishes, $\rmpow_\cU=0$,  and the exponent of the 
 second Symanzik polynomial is given
  by the charge, $\rmpow_\cF = Q_\cF$. Examples with this exponent structure can be obtained for example for $D=4$, see the next section. 
  For such graphs the integrand $\om_Q = \si/\cF^Q$ leads to the differential forms that enter the residue construction of the cohomology 
  of special Fano varieties. The maximally cut integrals $I_{\rm mcut}$ thus take the form 
   \beq
  I_{\rm mcut}~=~  \int_T \frac{\si}{\cF^Q},
  \lleq{mcut-integrals}
  which define period integrals of the special Fano variety via (\ref{Omega-integral}).

\vskip .3truein

 \section{Feynman integrals of special Fano type}
 
As mentioned above already, there are many Feynman graphs for which the parametric representation following Feynman and Symanzik leads to 
hypersurfaces of special Fano type. In this section examples are considered that arise from processes in the standard model 
after tensor reductions from graphs involving different types of particles.  

As a first illustration of a Feynman graphs of special Fano type we can consider the two-loop vertex corrections shown in Fig. 1, 
 graphs that arise via tensor-spinor reduction from Feynman integrals in different standard model sectors.
Historically, such graphs were considered for example in the early days of QED in the computations of the electron 
magnetic moment \cite{kk49, p57, s57}. 
 In more recent years such graphs have in particular been of interest in the context of QCD \cite{g83, vn85} and  
 Higgs induced physics \cite{f04etal, a04etal}. The first graph in Fig. 1 will in particular be shown below to be part 
 of an infinite sequence of Feynman integrals of special Fano type.
   \begin{center}
  \begin{picture}(250,40)
  \put(0,20){\line(1,0){10}}
  \put(10,20){\line(2,1){30}}
  \put(10,20){\line(2,-1){30}}
  \put(25,12){\line(0,1){15}}
  \put(33,9){\line(0,1){22}}
  
  \put(60,20){\line(1,0){10}}
  \put(70,20){\line(2,1){11}}
  \put(85,27){\circle{8}}
  \put(90,29){\line(2,1){10}}
  \put(70,20){\line(2,-1){30}}
  \put(96,8){\line(0,1){23}}
  
  \put(120,20){\line(1,0){10}}
  \put(130,20){\line(2,1){30}}
  \put(130,20){\line(2,-1){30}}
  \put(153,9){\line(0,1){8}}
  \put(153,20){\circle{7}}
  \put(153,25){\line(0,1){7}}
  
  \put(180,20){\line(1,0){10}}
  \put(190,20){\line(2,1){30}}
  \put(190,20){\line(2,-1){30}}
  \put(212,9){\line(0,1){22}}
   \put(212,16){\line(-1,1){10}}

 \put(240,20){\line(1,0){10}}
  \put(250,20){\line(2,1){38}}
  \put(250,20){\line(2,-1){38}}
  \put(284,4){\line(-1,2){13}}
  \put(269,11){\line(1,2){5}}
  \put(276,24){\line(1,2){6}}
  \end{picture}
  \end{center}
  \begin{quote}
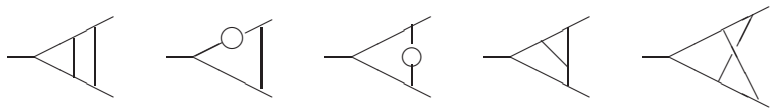
  {\bf Fig. 1} ~The basic set of five graphs associated to the two-loop contribution to magnetic moment of the electron in QED after scalar reduction.
  \end{quote}

 The Symanzik polynomials of the two-loop three-point functions in Fig. 1 all lead to the same 
 geometric configuration. They have the characteristics $(\ell, n_\rmint, n_\rmext) = (2, 6, 3)$ and the expansion 
 in (\ref{expansion})  leads 
 for the six internal momenta $q_a, a = 1,..., 6$ and the two independent momenta $k_j, j=1,2$ to the matrix $M$ that 
  describes the quadratic dependence on $k_j$, which leads to the first Symanzik polynomial  $\cU = \rmdet~ M$.
  This polynomial describes   a quadratic hypersurface $X_\cU$ in ordinary project space $X_\cU \in \mathP_5[2]$,  which is 
  rigid in the sense that it is independent of the physical parameters.
  The second Symanzik polynomial is of degree three, 
  leading to a cubic fourfold $X_\cF \in \mathP_5[3]$ that depends on the kinematic invariants and the masses. 
  These hypersurfaces have first Chern classes $c_1(X_\cU) = 4h$ and $c_1(\cF)=3h$, respectively, hence define 
   special Fano varieties with $Q_\cU=3$ and $Q_\cF=2$, thus providing the  
  building blocks of the above Feynman integral, depending on the spacetime dimension. 
  With the charges $Q_\cU=3$ the exponent of $\cU$ of a 2-loop integral with $\nu_a=1$ leads to 
  $(Q_\cU \ell - (\ell+1)D/2)$, which  vanishes for $D=4$. Since $Q_\cF=2$ the exponent of $\cF$ in eq. (\ref{pow-F})
  is given by $\rmpow_\cF = Q_\cF$. This shows that the Feynman integrand for this graph is precisely of the form 
  needed to define the form $\Om \in H^{3,1}(X_\cF)$ as in eq. (\ref{Omega-integral}). For graphs of this type the maximally 
  cut Feynman integral with $\nu_a=1$ is of the form of the period integral (\ref{mcut-integrals}).
  
 The cohomology of the generic fourfolds in the configuration $\mathP_5[3]$ can be summarized by the Hodge diamond given as
 \beq
  h^{p,q} ~=~ \delta^{p,q}, ~~~~p+q ~<~ 4
  \eeq
  and 
  \beq
  h^{p,q},~ p+q=4: ~~~~~0~~~~1~~~~21 ~~~~1 ~~~~0,
  \eeq
   where the remaining Hodge numbers are determined by symmetry.
  Because this fourfold is of special Fano type the Hodge numbers $h^{p,0}$ vanish for $p>0$.  The groups $H^{i,i}, i < 2$ have dimension one via 
 Lefschetz' theorem. Since $Q=2$ the Hodge decomposition of the intermediate cohomology
  contains the groups $H^{3,1} \cong H^{1,3}$, which have dimension one, as well as $H^{2,2}$. 
   This leads to a unique $(3,1)$-form which is the analog of the holomorphic $(n,0)$-form that exists in 
   Calabi-Yau manifolds of complex dimension $\rmdim_\mathC X=n$.
   A generic one-parameter family of these varieties thus leads to a Picard-Fuchs equation of order three, the same order that 
   is obtained from K3 surfaces.
    The link between fibers in the configuration $\mathP_5[3]$ and K3 surfaces was considered in a different context 
    in ref. \cite{kls08}, where it was shown
    that the $\Om$-motive of the fourfold leads to the same modular form as the corresponding K3 fiber. This shows that 
    the same motive can appear 
    in quite different varieties, even having different dimensions and being of different type. 
    A similar phenomenon was found in ref. \cite{kls10} where it was shown that the $\Om$-motives of 
    both smooth and singular varieties can lead to the same modular forms, indicating a degeneration of the Hodge structure 
    at singular fibers in a 
    family, which in turn affects the order of the Picard-Fuchs operator. More recently a similar embedding of a motive in 
    two different configurations among 
    CY threefolds  was described in ref. \cite{b21etal}.
    
 The special Fano varieties defined by the family of cubic fourfolds have more recently become of  interest, in particular their relation to 
 K3 surfaces, and more detailed information has become available that was not discussed in \cite{rs92, rs94, cdp93}. 
 This includes the determination of the intersection matrix $\cC$, which determines the K\"ahler 
 potential of the moduli space of complex deformations via the periods, as in eq. (\ref{kaehler-pot-cs2}).  The lattice $H^4(X,\mathZ)$ is 
 a unimodular lattice of signature $(21,2)$ relative to the intersection matrix $\cC$, which on the primitive part is given by \cite{h97}
 \beq
  \cC_\rmprim ~=~ E_8^{\oplus 2} \oplus U^{\oplus 2} \oplus A_2,
  \eeq
  where $E_8$ denotes the intersection matrix of the group $E_8$ by abuse of notation, while 
  \beq
  U ~=~ \left(\matrix{ 0 &1 \cr 1 &0\cr}\right), ~~~~~A_2 ~=~ \left(\matrix{2 &1\cr 1 &2\cr}\right)
  \eeq
  denote the hyperbolic plane and the special case $n=2$ for the $A_n$ diagrams respectively.
  The matrix $\cC$ is very closely related to the intersection matrix 
 of K3 surfaces \cite{bpv84}, which contains the same building blocks $E_8$ and $U$, but replaces the $A_2$ factor by another hyperbolic matrix. 
 
 The period domain associated to this cohomology can be viewed as a hypersurface in the projective space of the complexified lattice, 
 leading to a hermitian symmetric domain. The monodromy group of cubic fourfolds is given by the subgroup of the orthogonal 
 group $O(H^4(X, \mathZ))$ defined by the invariance of the hyperplane induced element $h^2$ and the constraint that the spinor norm is unity.
 
 The relation between the loop number and dimension of the variety is different in the case of special Fano varieties as compared to the CY cases 
 considered in the literature 
         \cite{ b09, bs10, mwz11, mwz12, bv13, bkv14, bkv16, fs18, b18etal, b18etal-b, bf19etal, b19etal, k19etal, vv20, b20etal, f21etal,  b21etal, 
            fh22, d22etal,  pww22, dhw22, pww22b, d22etal-b, d23etal, c23etal, g23etal, mh23, dk23etal, m23etal, d24etal, m24etal, j24etal, 
            d24etal-b, fnt24}.  
  For the case of $\ell-$loop banana graphs the number of internal momenta is $\ell+1$ and hence the ambient space has 
 dimension $\ell$ leading to associated hypersurfaces $X_b$ of dimension $\rmdim_\mathC X_b = (\ell-1)$. 
 The same holds for the traintrack diagrams considered in \cite{b18etal-b}. In the present case the special Fano variety associated 
 to a two-loop diagram is of dimension four. 
 It is in this context of interest that if one subtracts from the intermediate cohomology  the one ambient space induced cohomology 
 from that of the cubic hypersurface the intermediate cohomology agrees with that of  K3 surfaces.
  Thus the motive obtained from this special Fano variety is of K3 type.
 
 As mentioned above, there are also graphs for which the variety $X_\cF$ determined by the second Symanzik polynomial is of special Fano type, 
 while the variety $X_\cU$ is not.  This is analogous  to the $\ell$-loop banana graphs, for which $X_\cF$ is of CY type but $X_\cU$ is not. 
 Prominent examples of such special Fano type Feynman integrals include at the two-loop level the six-point function shown in Fig. 2, and at the 
 three-loop level the propagator graphs shown in Fig. 3.
 
  \begin{center}
 \begin{picture}(150, 70)
 \put(10,30){\line(1,0){20}}
 \put(30,30){\line(1,1){20}}
 \put(30,30){\line(1,-1){20}}
 
 \put(50,50){\line(2,-1){20}}
 \put(50,10){\line(2,1){20}}
 \put(50,50){\line(0,1){15}}
 \put(50,10){\line(0,-1){15}}
 
 \put(70,20){\line(0,1){20}}
 \put(70,20){\line(2,-1){20}}
 \put(70,40){\line(2,1){20}}

 \put(90,50){\line(1,-1){20}}
 \put(90,10){\line(1,1){20}}
 \put(90,50){\line(0,1){15}}
 \put(90,10){\line(0,-1){15}}
 \put(110,30){\line(1,0){20}}
 \end{picture}
 \end{center}
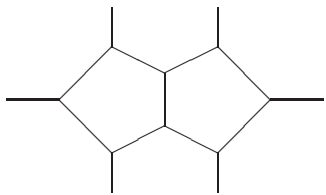
 \centerline{ {\bf Fig. 2} The double-pentagon graph with configuration $\mathP_8[3]$.}

 The two-loop six-point function in Fig. 2 has received attention for example in 
 \cite{b08etal}.  It is seen here to lead to a family of special Fano varieties 
 $X_\cF$ of charge $Q=3$ that live in the  configuration $X_\cF \in \mathP_8[3]$. Generic elements have cohomology 
 \beq
  h^{p,q}(X_\cF) ~=~ \delta^{p,q}, ~~~~p+q~<~ 7
  \eeq
  while the intermediate cohomology decomposes as 
  \beq
 \mathP_8[3], ~~  h^{p,q}, ~~p+q=7:~~~~~0 ~~0 ~~1~~84 ~~84 ~~1 ~~0~~0.
   \eeq
  The monodromy group here is given by the subgroup $\rmSpO(H^7(X, \mathZ), \vphi_H)$ of the symplectic group, 
  where $\vphi_H$ is the mod 2 quadratic refinement of the intersection 
  matrix. This configuration has been of interest in its string theory incarnation as the mirror configuration of 
  rigid Calabi-Yau spaces   \cite{rs92, rs94, cdp93}.
  Other Feynman integrals that lead to varieties in this configuration include the 2-loop hexagon-box, considered for example in 
   \cite{b08etal, k10etal, h21etal}.
  
 The families $X_\cF$ defined by the second Symanzik polynomial $\cF$ of the three-loop diagrams in Fig. 3 satisfy the special 
 Fano quantization condition with charge $Q=2$. 
  \begin{center}
  \begin{picture}(370,60)
   \put(10,30){\line(1,0){20}}
   \put(90,30){\line(1,0){20}}
   \put(60,30){\oval(60,30)}
   \put(50,45){\line(0,-1){30}}
   \put(70,45){\line(0,-1){30}}
  
   \put(130,30){\line(1,0){20}}
   \put(210,30){\line(1,0){20}}
   \put(180,30){\oval(60,30)}
    \put(170,45){\line(2,-3){20}}
   \put(190,45){\line(-2,-3){20}}
    
 \put(250,30){\line(1,0){20}}
 \put(290,30){\circle{40}}
 \put(290,50){\line(0,-1){20}}
 \put(290,30){\line(1,-1){14}}
 \put(290,30){\line(-1,-1){14}}
 \put(310,30){\line(1,0){20}}

 \end{picture}
  \end{center}
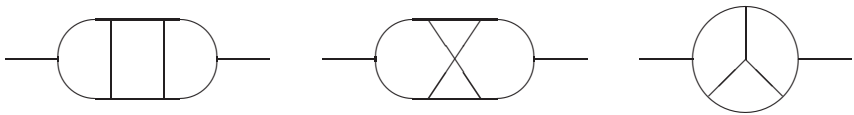
\centerline{{\bf Fig. 3}~ Three-loop propagator graphs in $\phi^3$.}

 These hypersurfaces of dimension six define families  in the configuration space of quartic sixfolds $X_{\cF_i} \in \mathP_7[4]$. Generic elements 
 of this configuration have Hodge numbers below the middle dimension given by 
  \beq
   h^{p,q} ~=~ \delta^{p,q}, ~~~~p+q<6
   \eeq
   while the intermediate cohomology decomposes as 
   \beq
    h^{p,q}, ~p+q=6:~~~~~0~~~1~~~266~~~1108~~~266~~~1~~~0.
    \eeq
    Three loop graphs of this and similar type have recently been of interest in the self-energy computations in QED for both the 
    electron \cite{d24etal} and the photon \cite{fnt24}.

There exist also infinite sequences of Feynman graphs that lead to Fano varieties of special type.  
 A first example of such a sequence of  graphs can be constructed by considering iterated Mercedes-Benz graphs with 
 increasing loop number. Each iteration is defined here by adding a circle, thereby adding six internal lines. 
 The lowest order graphs in this sequence are illustrated in Fig. 4. 
Starting with the second Symanzik polynomial of the three-loop Mercedes-Benz graph in Fig. 3, the first iteration at six loops leads to  the special 
Fano variety in the configuration $X_\cF \in \mathP_{13}[7]$, shown in the left panel of Fig. 4. With each  iteration the number of 
loops increases  by three, leading to a loop number $\ell_n = 3+3n$. The charge $Q$ of all the varieties constructed in this way is $Q=2$, 
hence the configuration determined in this way is given by $\mathP_{7+6n}[4+3n]$.
    \begin{center}
\begin{picture}(200,50)
\put(50,20){\circle{25}} 
\put(50,20){\circle{40}}
\put(50,20){\line(0,1){19}}
\put(50,20){\line(1,-1){14}}
\put(50,20){\line(-1,-1){14}}
\put(10,20){\line(1,0){20}}
\put(70,20){\line(1,0){20}}

\put(160,20){\circle{20}} 
\put(160,20){\circle{30}}
\put(160,20){\circle{40}}
\put(160,20){\line(0,1){20}}
\put(160,20){\line(1,-1){14}}
\put(160,20){\line(-1,-1){14}}
\put(125,20){\line(1,0){15}}
\put(180,20){\line(1,0){19}}
\end{picture}
\end{center}
\begin{quote}
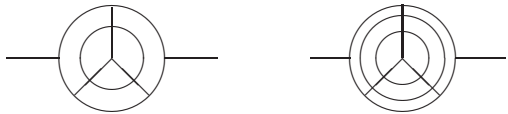
{\bf Fig. 4} ~The first two iterations of the  3-loop Mercedes-Benz graph with $n=1,2$. Increasing $n$ leads to an infinite 
sequence MB$_\infty$ of varieties of special Fano type.
\end{quote}

It is also of interest to consider higher $n$-point functions, which are of relevance for example in jet physics.  
This has led to an extensive literature already 
for the one-loop case, see for example \cite{bdk93} and follow-up papers for phenomenological aspects. 
Such 1-loop $n$-point functions lead an 
infinite sequence of graphs with special Fano type varieties for both $\cU$ and $\cF$ in even dimensions.

An extension that leads to a double infinite sequence of special Fano type of integrals can be constructed by extending an 
$n$-point loop by 1-loop boxes, leading to an infinite number of infinite sequences of Feynman integrals, illustrated  in Fig. 5.
\begin{center}
   \begin{picture}(50,60)
   \put(-40,20){\line(-1,-1){13}}
   \put(-40,40){\line(-1,1){13}}
      
   \put(-40,20){\line(0,1){20}}
   \put(-40,20){\line(1,0){20}}
   \put(-40,40){\line(1,0){20}}
   \put(-20,20){\line(0,1){20}}
   
   \put(-20,20){$\dots$}
   \put(-20,39){$\dots$}
  
   \put(6,20){$\dots$}
   \put(6,39){$\dots$}
   
   \put(20,20){\line(0,1){20}}
   \put(20,20){\line(1,0){20}}
   \put(20,40){\line(1,0){20}}
   \put(40,20){\line(0,1){20}}

   \put(40,20){\line(2,-1){15}}
   \put(40,40){\line(2,1){15}}
   \put(55,12){\line(2,1){15}}
   \put(55,48){\line(2,-1){15}}
   
   \put(70,40){\line(1,-2){3}}
   \put(68,31){$\vdots$}
   \put(68,20){$\vdots$}
   \put(70,19){\line(1,2){3}}
  
   \put(70,19){\line(1,-1){13}}
   \put(70,41){\line(1,1){13}}
   
   \put(55,12){\line(0,-1){12}}
   \put(55,48){\line(0,1){12}}
   
   \end{picture}
   \end{center}
\centerline{{\bf Fig. 5} ~A doubly infinite sequence of graphs.}

This sequence of graphs can be characterized by the number $n_\rmpoly=(p-1)$ of lines in the 
1-loop $n$-point graph, the total number $\ell$ of loops, and the charge $Q$ of the associated 2nd Symanzik polynomial. 
The special Fano quantization condition then is given by 
\beq
 n_\rmpoly ~-~ Q(\ell+1) + 3\ell - 1 ~=~ 0.
 \eeq
 Depending on the charge $Q$ and the number of external lines these graphs can lead to an infinite number of graphs of special Fano type.
Consider for example an end-loop with four external momenta $p_i$. By adding an arbitrary number of one-loop boxes 
we obtain an infinite sequence of varieties $X_\cF$ of special Fano type varieties with charge $Q=3$, starting with the 
hexagon-box graph, which lives in the configuration $\mathP_8[3]$. This configuration was encountered above in the context 
of the double-pentagon graph, see Fig. 2.
 The hexagon-box graph has recently received some attention in a different context, see for example \cite{h21etal}.

 Finally, an example for which the first Symanzik polynomial leads to a hypersurface of special Fano type, while $X_\cF$ does not, 
 is given by the 2-loop vertex graph  in the quartic theory. 
 The variety $X_\cU$ in this case has charge $Q=2$ and is a complex surface  of the configuration $X_\cU \in \mathP_3[2]$.
 
 Aside from ubiquitous applications in the particle physics of the standard model, Feynman integrals have more recently 
 been useful in gravitational physics. 
 While the application of quantum field theoretic methods to obtain classical limits is a  topic with deep roots going 
 back half a century, see e.g. \cite{i71}, it has 
 been revived more recently in the wake of the LIGO discovery to compute the effects of black hole interactions. 
 Using the framework of two massive
 scalar fields coupled to gravity the classical potential has in particular been computed in the PM expansion to 
 4th order \cite{crs18, bern19etal-b, bern21etal}.
 In this context one encounters Feynman integrals that are of special Fano type.

\vskip .3truein

\section{Conclusion}

In this paper the framework of geometric analyses of Feynman integrals has been extended by considering a class of manifolds that was 
originally considered in the context of Landau-Ginzburg string vacua and  mirror symmetry. These special Fano varieties 
 are characterized by a nontrivial first Chern class that is integral in terms of a charge $Q$ associated to the Landau-Ginzburg theory. 
 As a consequence they have a cohomology structure that is reduced compared to that of CYs, with a Hodge decomposition  
 that depends not only on the dimension of the variety, but also on the charge $Q$. This has a number of implications, 
 for example a reduced nilpotency exponent for the monodromy of varieties of special Fano type that affects the structure 
 of the period equation.

In the context of quantum field theory these varieties can describe the geometry induced by Feynman integrals of the standard model 
via  the Symanzik polynomials $\cU$ and $\cF$ obtained after tensor-spinor reduction.  Particularly 
interesting are Feynman graphs for which both $\cU$ and $\cF$  are of special Fano type, but graphs 
for which either $\cU$ or $\cF$ are of this type 
are also of interest. The existence of the higher-dimensional counterpart of the holomorphic $n$-form of critical manifolds 
leads to a relation between the physical parameter space of Feynman integrals, spanned by the kinematic invariants and the masses,  
and  the moduli space as part of the cohomology of the special Fano variety. 
This generalizes the geometric analyses that have been used  so far in literature,  including in particular the deformation theory that leads to the 
Picard-Fuchs equation associated to the Feynman graphs, in a way 
that generalizes the period equations considered, for example in the class of banana graphs, in the CY oriented literature.

\vskip .2truein

{\bf Acknowledgement.} \\
It is a pleasure to thank Philip Candelas, Monika Lynker and Stephan Stolz for discussions and correspondence.

\end{document}